\documentclass[twocolumn,aps,amsmath,amssymb,prl,epsf,showpacs,preprintnumbers]
{revtex4}

\usepackage{psfig}
\usepackage{graphicx}
\usepackage{dcolumn}
\usepackage{bm}

\newcommand{\lp}{\ell_{\mathrm P}}
\newcommand{\be}{\begin{equation}}
\newcommand{\ee}{\end{equation}}
\newcommand{\bq}{\begin{eqnarray}}
\newcommand{\eq}{\end{eqnarray}}

\newcommand{\lra}{\longrightarrow}

\usepackage{psfig}
\usepackage{graphicx}
\usepackage{dcolumn}
\usepackage{bm}

\newcommand{\Ph}{\ensuremath{\Phi}}
\newcommand{\Pdot}{\ensuremath{\dot{\Phi}}}

\begin{document}
\preprint{IGPG-05/6-8}
\title{Quantum evaporation of a naked singularity}

\author{Rituparno Goswami$^\ast$, Pankaj S. Joshi$^\ast$, Parampreet 
Singh$^\dagger$}
\affiliation{$^\ast$Tata Institute for Fundamental Research, Colaba, 
Mumbai 400005, India}
\affiliation{$^\dagger$Institute for Gravitational Physics and Geometry, 
Pennsylvania State University, University Park, PA 16802, USA}

\begin{abstract}
We investigate here quantum effects in gravitational collapse of 
a scalar field model which classically leads to a naked singularity. 
We show that non-perturbative semi-classical modifications near the 
singularity, based on loop quantum gravity, give rise to a strong outward 
flux of energy. This leads to the dissolution of the collapsing cloud 
before the singularity can form. Quantum gravitational effects thus 
censor naked singularities by avoiding their formation. Further, 
quantum gravity induced mass flux has a distinct feature which may 
lead to a novel observable signature in astrophysical bursts.

\end{abstract}

\maketitle

Naked singularities are one of the most exotic objects 
predicted by classical general relativity. Unlike their black hole 
siblings, they can be in principle directly observed by an
external observer. There have been many investigations which  
show that given the initial density and pressure profiles for a matter 
cloud, there are classes of collapse evolutions that lead to naked
singularity formation (see e.g.  
\cite{review} 
for some recent reviews), subject to an energy condition and 
astrophysically reasonable equations of state such as dust, perfect
fluids and such others. This has led to extensive debates on their 
existence, with a popular idea being cosmic censorship conjectures which 
forbid classical nakedness 
\cite{CCC}. 
Since naked singularities originate in the regime where classical 
general relativity is expected to be replaced by quantum gravity, 
it has remained an outstanding problem whether a quantum theory of 
gravity resolves their formation. Also, with the lack of observable 
signatures from the Planck regime, naked singularities could in fact 
be a boon for a quantum theory of gravity. Because, the singularity 
being visible, any quantum gravitational signature originating in 
the ultra-high curvature regime near a classical singularity can 
in principle be observed, thus providing us a rare test for 
quantum gravity.

One of the non-perturbative quantizations of gravity is loop 
quantum gravity 
\cite{lqg_review} 
whose key predictions include Bekenstein-Hawking entropy formula 
\cite{bek_hawking}.
Its application to symmetry reduced mini-superspace quantization 
of homogeneous spacetimes is called loop quantum cosmology 
\cite{martin}
whose success includes resolution of the big bang singularity
\cite{bigbang}, 
initial conditions for inflation 
\cite{superinflation,inflationcmb}, 
and possible observable signatures in cosmic microwave background radiation 
\cite{inflationcmb}. 
These techniques have also been applied to resolve black hole singularity 
in a scalar field collapse scenario 
\cite{bhole}.

Since the dynamics of a generic collapse is very complicated and  
tools to address such a problem in quantum gravity are still under 
development, it is useful to work with a simple collapse scenario as of 
a scalar field. It serves as a good toy model to gain insights on 
the role of quantum gravity effects at the late stages of gravitational 
collapse. Existence of naked singularities in these models is well-known 
\cite{scalar} 
and one of the simplest setting is to consider an initial configuration 
of a  homogeneous and isotropic scalar field $\Phi = \Phi(t)$ with a 
potential $V(\Phi)$ (given by eq.(\ref{potential})) and the canonical 
momentum $P_\Phi$. In this case it has been shown that fate of the 
singularity being naked or covered depends on the rate of gravitational collapse 
\cite{JG1}. 
For an appropriately chosen potential, formation of trapped surfaces 
can be avoided even as the collapse progresses, resulting in a 
naked singularity with an outward energy flux, in principle observable. 
Since the interior of homogeneous scalar field collapse is described by 
a Friedmann-Robertson-Walker (FRW) metric, techniques of loop quantum cosmology can be used
to investigate the way quantum gravity modifies the collapse.

Let us consider the classical collapse of a homogeneous scalar 
field $\Phi(t)$ with potential $V(\Phi)$ and the canonical momentum  
for the  marginally bound $(k=0)$ case. The interior metric is 
given by
\begin{equation}
ds^2=-dt^2+a^2(t)\left[dr^2+r^2d\Omega^2\right] ~
\label{eq:FRW}
\end{equation}
with classical energy density and pressure of the scalar field,
\be
\rho(t)=\Pdot^2/2+V(\Ph), ~~ p(t)=\Pdot^2/2-V(\Ph) ~. \label{rhop}
\ee
The dynamical evolution of the system is obtained from the 
Einstein equations which yield 
\cite{JG1}
\begin{equation}
\dot{R}^2 R = F(t,r), \, \rho=F_{,r}/\kappa a R^2, \, 
p =  -\dot{F}/\kappa R^2\dot{R}
\label{eq:ein4}
\end{equation}
Here $\kappa = 8 \pi G$, and $F(t,r) = (\kappa/3) \rho(t) r^3 a^3$ 
has interpretation of the mass function of the collapsing 
cloud, with $F\ge0$ and $R(t,r)=ra(t)$ is the area
radius of a shell labeled by comoving coordinate $r$.  
In a continual collapse the area radius of a shell at a constant 
value of comoving radius $r$ decreases monotonically. 
The spacetime region is trapped or otherwise, depending on 
the value of mass function. If $F$ is greater (less) than $R$, 
the the region is trapped (untrapped). The boundary of the trapped 
region is given by $F=R$.

The collapsing interior can be matched at 
some suitable boundary $r=r_b$
to a generalized Vaidya exterior geometry, given as
\cite{wang}, 
\begin{equation}
ds^2=-(1-2M(r_v,v)/r_v)dv^2-2dvdr_v+r_v^2d\Omega^2 ~.
\label{eq:metric2}
\end{equation} 
The Israel-Darmois conditions then lead to 
\cite{JG1,wang} 
$r_ba(t)=r_v(v)$, $F(t,r_b)=2M(r_v,v)$ and
\begin{equation}
M(r_v,v)_{,r_v}\, = \, F/2r_ba \, + \, r_b^2a\ddot{a} ~.
\label{eq:match5}
\end{equation}

The form of the potential that leads to a naked singularity 
is determined as follows. The energy density of scalar field can
be  written in a generic form as $\rho = l^{n-4} a^{-n}$, where $n > 0$ 
and $l$ is a proportionality constant. Using energy conservation equation, 
this leads to the pressure $p = [(n-3)/3] ~ l^{n-4} a^{-n}$. 
On subsituting eq.(\ref{rhop}) in these we obtain 
\cite{JG1} 
\be \label{potential}
\Phi = - \sqrt{n/\kappa} \ln a, ~~ V(\Phi) = (1-n/6) 
l^{n-4} e^{\sqrt{\kappa n} ~\Phi} ~.
\ee
Then it is easily seen that $F/R = (\kappa/3) l^{n-4} a^{2-n} r^2$. 
Thus in the collapsing phase as $a \lra 0$, whether or not
the trapped surfaces form is determined by the value of $n$. 
It is  straightforward to check that for $0<n<2$, if 
no trapped surfaces exist initially then no trapped surfaces 
would form till the epoch $a(t) = 0$ 
\cite{JG1}, 
with $a(t)=\left(1-n \, t/2\sqrt{3}\right)^{2/n}$.
\label{eq:a1}

The absence of trapped surfaces is accompanied by a negative pressure 
implying that for a constant value of the comoving coordinate $r$, $\dot F$ 
is negative and so the mass contained in the cloud of that radius keeps  
decreasing. This leads to a classical outward energy flux. 
As the collapse proceeds, the scale factor vanishes in finite 
time and physical densities blow up, leading to a naked singularity. 
Since no trapped surfaces form during collapse, the outward energy flux 
shall in principle be observable. However, near the singularity when energy 
density is close to Planckian values, this classical picture has to be modified
and we need to investigate the scenario incorporating quantum gravity 
modifications to the classical dynamics.

Let us hence consider the non-perturbative semi-classical modifications 
based on loop quantum gravity for the interior. The underlying geometry 
for the FRW spacetime in loop quantum cosmology is discrete and both 
the scale factor and the inverse scale factor operators have
discrete eigenvalues 
\cite{Bohr}. 
In particular, there exists a critical scale $a_* = \sqrt{j \gamma/3} \lp$ 
below which the eigenvalues of the inverse scale factor become proportional 
to the positive powers of scale factor. Here $\gamma \approx 0.2375$ is 
the Barbero-Immirzi parameter 
\cite{bek_hawking}, 
$\lp$ is Planck length and $j$ is a half-integer free parameter 
which arises because inverse scale factor operator is computed by 
tracing over SU(2) holonomies in an irreducible spin $j$ representation.
The value of this parameter is arbitrary and shall be constrained 
by phenomenological considerations.

The change in behavior of the classical geometrical density ($1/a^3$)
for scales $a \lesssim a_*$, can be well approximated by 
\cite{superinflation}
\be
d_{j}(a) = D(q) \, a^{-3}, ~~ q := a^2/a_*^2, ~~ a_* := 
\sqrt{j \gamma/3} \, \lp \label{density}
\ee
with
\begin{eqnarray} \label{defD}
&&D(q) = \left( {8/ 77}\right)^6 q^{3/2} \Big\{7 \Big[(q+1)^{11/4}
-|q-1|^{11/4}\Big] \nonumber \\
&&~~~ -11q\Big[(q+1)^{7/4}-{\rm sgn}\,(q-1) |q-1|^{7/4}\Big]
\Big\}^6.\!\label{D}
\end{eqnarray}
For $a \ll a_*$, $d_j \propto (a/a_*)^{15} a^{-3}$ and for 
$a \gg a_*$ it behaves classically with $d_{j} \approx a^{-3}$. 
The scale at which transition in the behavior of the geometrical 
density takes place is determined by the parameter $j$.

At the fundamental level the dynamics in the loop quantum regime 
is discrete, however, recent invsetigations pertaining to the evolution 
of coherent states have shown that for scales 
$a_0 = \sqrt{\gamma} \lp  \lesssim a \lesssim a_* = \sqrt{j \gamma/3} \lp$, 
dynamics can be described by modifications
to Friedmann dynamics on a continuous spacetime 
\cite{time} with the modified matter Hamiltonian 
\be 
{\cal H}_\Phi = d_{j}(a)\,  P_\Phi^2/2 + a^3 \, 
V(\Phi) ~ \label{hamphi}
\ee
and the modified Friedmann equation 
\be \label{modfred}
\dot a^2/a^2 = (\kappa/3) (\dot \Phi^2/2 D + V(\Phi)) ~
\ee
which is obtained by the vanishing of the total Hamiltonian constraint 
and the Hamilton's equations: $\dot \Phi =  d_{j}(a) P_\Phi, ~~ \dot P_\Phi  
= - a^3 \, V_{,\Phi}(\Phi)$ 
\cite{superinflation}. 
These also lead to the modified Klein-Gordon equation
\be
\ddot \Phi + \left(3 \dot a/a - \dot D(q)/D(q) \right) 
\, \dot \Phi + 
D(q) \, V_{,\Phi}(\Phi) ~ = 0 ~. \label{kgeq}
\ee
Since at classical scales ($a \gg a_*$) $D \approx 1$, the 
modified dynamical equations reduce to the standard Friedmann dynamical 
equations. For scales $a \lesssim a_*$, the $\dot \Phi$ term acts like a 
frictional term for a collapsing phase. We note that since 
semi-classical modifications for inhomogeneous case are still not known,
we cannot do a complete quantum analysis of interior and exterior.
The exterior is assumed to remain classical. Further, as a continuous
spacetime can be approximated till scale factor $a_0$, the matching 
of interior and exterior spacetimes remains valid during the semi-classical
evolution.

The modified energy density and pressure of the scalar field in 
the semi-classical regime can be similarly obtained from the eigenvalues 
of density operator and using the stress-energy 
conservation equation 
\cite{density} 
\be
\rho_{\rm eff} = d_{j}(a) \, {\cal H}_\Phi = \dot \Phi^2/2 
+ D(q) \, V(\Phi) 
\label{energyden1}
\ee
and
\be
p _{\rm eff}= \bigg[1 - \frac{2}{3} \, \frac{1}{(\dot a/a)} \, 
\frac{\dot D(q)}{D(q)}
\bigg] \, \frac{\dot \Phi^2}{2} - D(q)\, V(\Phi) - \frac{\dot D(q)}
{3 (\dot a/a)} \, V(\Phi) ~.
\ee
It is then straightforward to check that $p_{\rm eff}$ is 
generically negative for $a \lesssim a_*$ and for $a \ll a_*$ it becomes 
very strong. For example, at
$a \sim a_0$, $p_{\rm eff} \approx - 9 \rho_{\rm eff}$. 
This is much stronger than
its classical counterpart $p = [(n-3)/3] ~ \rho$ with $0 < n < 2$.
Thus we expect a strong burst of outward energy flux in the 
semi-classical regime.
Further, for $a \ll a_*$, $D(q) \ll 1$ and the Klein-Gordon 
equation yields $\dot{\Phi}\propto a^{12}$. Hence from the eq. 
(\ref{energyden1})
we easily see that the effective density, instead of blowing up, 
becomes extremely small and remains finite. 

The modified mass function of the 
collapsing cloud can be evaluated using eq.(\ref{eq:ein4}) and eq.(\ref{modfred}), 
\be
F = (\kappa/3) (d_{j}^{-1} \, \dot \Phi^2/2 
+ a^3 \, V(\Phi)) \, r^3 ~.
\ee
In the regime $a \sim a_0$, $d_{j}^{-1} \dot \Phi^2$ becomes proportional 
to $a^{12}$, the potential term becomes 
negligible 
and thus the mass function becomes vanishingly small at small scale factors.

\begin{figure}
\begin{center}
\includegraphics[width=8cm,height=6cm]{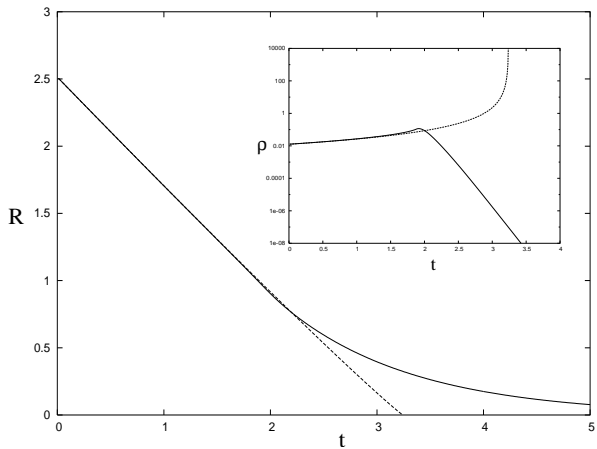}
\end{center}
\vskip-0.5cm \caption{Evolution of area radius with time. The 
classical evolution (dashed) leads to naked singularity
in finite time whereas in semi-classical evolution (solid) it is avoided. 
Inset shows evolution of energy density (in Planck units) with time. The
parameters chosen are $n = 1.9$ and $j = 100$. } \label{fig1}
\end{figure}

The picture emerging from loop quantum modifications 
to collapse is thus following.

$\bullet$ Before the area radius of
the collapsing shell reaches $R_* = r a_*$ at $t = t_*$, collapse 
proceeds as per classical dynamics and as smaller scale 
factors are approached $\dot \Phi$ and the energy density $\rho 
\propto a^{-n}$ increase. The mass function is proportional to $a^{n-3}$
and (as $0 < n < 2$) it decreases with decreasing scale factor so
there is a mass loss to the exterior, which is also understood 
from existence of negative classical pressure. 

$\bullet$ As the collapsing cloud reaches $R_*$, the geometric 
density classically given by $a^{-3}$, modifies to $d_j$ and the dynamics 
is governed by the modified Friedmann and Klein-Gordon equations. The 
scalar field which experienced anti-friction in classical regime, now 
experiences friction leading to decrease of $\dot \Phi$.

$\bullet$ The slowing down of $\Phi$ 
decreases the rate of collapse and formation of singularity is delayed. 
Eventually when scale factor becomes smaller than $a_0$ this leads to 
breakdown of continuum spacetime approximation and semi-classical dynamics. 
Discrete quantum geometry emerges at this scale 
\cite{time} 
and the dynamics can only be  described by quantum difference equation. 
The naked singularity is thus avoided till the scale factor at which a  
continuous spacetime exists.

We show the evolution of area radius in time as 
collapse proceeds in Fig.\ref{fig1}. The semi-classical 
evolution (solid curve)  closely follows classical trajectory 
(dashed) till the time $t_*$. Within a finite time after $t_*$, 
the classical collapse leads to a vanishing $R$ and naked 
singularity. However, the area radius never vanishes in the loop modified 
semi-classical dynamics and the naked singularity does not form as long
as the continuum spacetime approximation holds. 
The inset of Fig.\ref{fig1} shows the evolution of energy density 
in Planck units. Classical energy density (dashed curve) blows up 
whereas it remains finite and in fact decreases in the 
semi-classical regime.

\begin{figure}
\begin{center}
\includegraphics[width=8cm,height=6cm]{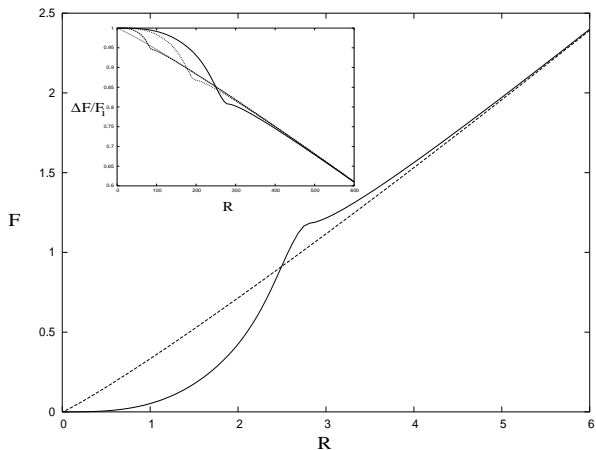}
\end{center}
\vskip-0.5cm \caption{Evolution of mass function with area radius 
for same parameters as in Fig.\ref{fig1}. Loop quantum evolution
(solid) leads to dissolution of all the mass of the collapsing shell. 
Dashed curve shows classical trajectory. Inset shows mass loss profile 
for $j = 10^6$ (outer), $j = 5.0\times10^5$ (middle)
and $j = 10^5$ (inner).  } \label{fig2}
\end{figure}

The phenomena of delay and avoidance of the naked singularity 
in continuous spacetime is accompanied by a burst of matter to the 
exterior. If the mass function at scales $a \gg a_*$ is $F_i$ and its 
difference with mass of the cloud for $a<a_*$ is $\Delta F = F_i - F$, 
then the mass loss can be computed as
\be
\frac{\Delta F}{F(a_i)}=\left[1-\frac{\rho_{\rm eff} d_j^{-1}}
{l^{n-4} a_i^{3-n}}\right] ~.
\ee
For $a < a_*$, as the scale factor decreases,  
the energy density and mass in the interior decrease and the negative 
pressure strongly increases. This leads to a strong burst
of matter.  The absence of trapped surfaces enables the quantum 
gravity induced burst to propagate via the generalized Vaidya exterior 
to an observer at infinity. 
The evolution 
of mass function is shown in Fig.\ref{fig2}. 
In the semi-classical regime,  ${\Delta F}/F_i$ approaches unity 
very rapidly. This feature is independent of the choice of
parameter $j$. The choice of potential causes mass loss to exterior 
in classical collapse also, but it is much smaller and in any case 
the classical description cannot be trusted at energy density greater 
than Planck, when we must consider quantum effects as above. 

Interestingly, for a given collapsing configuration, the scale 
at which the strong outward flux initiates depends on the loop parameter 
$j$ which controls $a_*$. If $j$ is large then burst occurs at an earlier 
area radius and vice versa. The inset of Fig.\ref{fig2} shows the 
mass loss ratio for different values of $j$. For all choices, 
${\Delta F}/F_i \rightarrow 1$, but the outgoing flux profile changes. 
The loop quantum burst has a distinct signature, at $a \sim a_*$ the 
flux decreases for a short period and then rapidly increases. Since the 
causal structure of classical spacetime is such that trapped surface 
formation is avoided, this quantum gravitational signature can be in 
principle observed by an external observer as a slight dimming and 
subsequent brightening of the collapsing star. This peculiar phenomena is 
directly related to the peak in the function $d_j(a)$, and depends 
solely on the value of parameter $j$. If we compare this to other 
phenomenological applications 
\cite{superinflation,inflationcmb,bhole}, 
this effect could not be masked by the role of other loop quantum 
parameters in a more general setting. This phenomena is thus a direct 
probe to measure $j$ and an observer can estimate the loop quantum 
parameter $j$ by observing the flux profile of the burst based on this 
mechanism and measuring the variation in luminosity of the 
collapsing cloud.

During such a burst most of the mass is ejected and this 
may dissolve the singularity. Thus non-perturbative semi-classical 
modifications may not allow formation of naked singularity as the 
collapsing cloud evaporates away due to super-negative pressures in 
the late regime. It has been demonstrated that these super-negative 
pressures would exist for arbitrary matter configurations 
\cite{density} 
which implies that results obtained here would hold even in a 
more general setting 
\cite{JG}. 
Loop quantum effects then imply a quantum gravitational 
cosmic censorship, alleviating the naked singularity problem. We 
note that the semi-classical effects do not show that the singularity 
is absent, it is only avoided till scale factor $a_0$, below which 
the semi-classical dynamics and matching may break down. If for a 
given choice of initial data, semi-classical dynamics is unable 
to completely dissolve the singularity, the final fate of naked 
singularity must be decided by using full quantum evolution. Even in 
such cases we have valuable insights from semi-classical loop quantum 
effects with the possibility of phenomenologically constraining 
the $j$ parameter.

In the toy model considered, we showed that the classical 
outcome and evolution of collapse is radically altered by the 
non-perturbative modifications to the dynamics. Our considerations 
are of course within the mini-superspace setting, and the general 
case of inhomogeneities and anisotropies remains open. However, the 
possibility of such observable signatures in astrophysical bursts, 
as originating from quantum gravity regime near singularity is 
intriguing, indicating that gravitational collapse scenario 
can be used as probes to test quantum gravity models.

{\it Acknowledgments:} We thank A. Ashtekar, M. Bojowald and R. Maartens 
for useful comments. PSJ thanks Bharat Kapadia for various discussions. PS is supported by Eberly research funds of Penn State 
and NSF grant PHY-00-90091.


\begin{thebibliography}{10}

\bibitem{review} A. Krolak, Prog. Theor. Phys. Suppl. {\bf 136}, 45 (1999);
P. S. Joshi, Pramana {\bf 55}, 529 (2000); T. Harada, H. Iguchi,
and K. Nakao, Prog.Theor.Phys. 107 (2002) 449.
  

\bibitem{CCC} R. Penrose, Riv. Nuovo Cimento, Num. Sp. 1, 252 (1969).

\bibitem{lqg_review} See for eg.,
 A.~Ashtekar and J.~Lewandowski, Class. Quant. Grav. {\bf 21} R53 (2004).


\bibitem{bek_hawking}
A.~Ashtekar et al.,
Phys. Rev. Lett. {\bf 80}, 904 (1998).

\bibitem{martin} M. Bojowald, gr-qc/0505057.



\bibitem{bigbang} M.~Bojowald, Phys.\ Rev.\ Lett.\ {\bf 86}, 5227 (2001).


\bibitem{superinflation}
M. Bojowald, Phys.\ Rev.\ Lett. {\bf 89},  261301  (2002).



\bibitem{inflationcmb} S. Tsujikawa, P. Singh, \& R. Maartens,  
Class. Quant. Grav. {\bf 21}, 5767 (2004).




\bibitem{bhole} M. Bojowald et al., Phys. Rev. Lett. {\bf 95} (2005) 091302. 

\bibitem{scalar} D. Chistodoulou, Ann. of Maths. {\bf 140}, 607 (1994);
{\bf 149}, 183 (1999); M. W. Choptuik, Phys. Rev. Lett. {\bf 70}, 9 (1993); 
A. M. Abrahams
and C. R. Evans, Phys. Rev. Lett. {\bf 70}, 2980 (1993); M. D. Roberts, Gen.
Relat. Grav. {\bf 21}, 907 (1989); J. Traschen,
Phys. Rev. D {\bf 50}, 7144 (1994); P. R. Brady, Phys. Rev. D {\bf 51}, 
4168 (1995); C. Gundlach, Phys. Rev. Lett. {\bf 75}, 3214 (1995).


\bibitem{JG1} P. S. Joshi \& R. Goswami, gr-qc/0410144; R. Giambo gr-qc/0501013.

\bibitem{wang} A. Wang \& Y. Wu, Gen. Rel. Grav. {\bf 31} (1), 107 (1999); 
P. S. Joshi \& I. H. Dwivedi, Class.Quant.Grav. {\bf 16}, 41 (1999).



\bibitem{Bohr} A.~Ashtekar, M.~Bojowald \& J.~Lewandowski, Adv. Theor. Math. 
Phys. {\bf 7}, 233 (2003).




\bibitem{time} M. Bojowald, P. Singh \&  A. Skirzewski,  Phys.Rev. D{\bf 70}, 083517 (2004); P. Singh, K. Vandersloot, 
Phys. Rev. D{\bf 72}, 084004 (2005).




\bibitem{density} P. Singh, Class. Quant. Grav. {\bf 22}, 4203 (2005).


\bibitem{JG} P.S.Joshi, I.H.Dwivedi, Phys.Rev. D {\bf 47}, 5357 (1993);  
P. S. Joshi, R. Goswami, Phys.Rev. D {\bf 69}, 064027 (2004).





\end{thebibliography}
\end{document}